\documentclass[onecolumn]{aastex6}
\usepackage{multirow}
\usepackage{color}

\newcommand{\kms}{\rm \,km\,s^{-1}}

\fullcollaborationName{The Friends of AASTeX Collaboration}

\begin{document}


\title{A massive molecular outflow in the dense dust core AGAL\,G337.916-00.477}


\author{Kazufumi Torii\altaffilmark{1}, Yusuke Hattori\altaffilmark{2}, Keisuke Hasegawa\altaffilmark{2}, Akio Ohama\altaffilmark{2}, Hiroaki Yamamoto\altaffilmark{2}, Kengo Tachihara\altaffilmark{2}, Kazuki Tokuda\altaffilmark{3, 4}, Toshikazu Onishi\altaffilmark{3}, Yasuki Hattori\altaffilmark{2}, Daisuke Ishihara\altaffilmark{2}, Hidehiro Kaneda\altaffilmark{2}, and Yasuo Fukui\altaffilmark{2}}

\altaffiltext{1}{Nobeyama Radio Observatory, 462-2 Nobeyama Minamimaki-mura, Minamisaku-gun, Nagano 384-1305, Japan}
\altaffiltext{2}{Department of Physics, Nagoya University, Chikusa-ku, Nagoya 464-8602, Japan}
\altaffiltext{3}{Department of Physical Science, Graduate School of Science, Osaka Prefecture University, 1-1 Gakuen-cho, Naka-ku, Sakai, Osaka 599-8531, Japan}
\altaffiltext{4}{Chile Observatory, National Astronomical Observatory of Japan, National Institutes of Natural Science, 2-21-1 Osawa, Mitaka, Tokyo 181-8588, Japan}

\begin{abstract}
Massive molecular outflows erupting from high-mass young stellar objects provide important clues to understanding the mechanism of high-mass star formation.
Based on new CO $J$=3--2 and $J$=1--0 observations using the Atacama Submillimeter Telescope Experiment (ASTE) and Mopra telescope facilities, we discovered a massive bipolar outflow associated with the dense dust core AGAL\,G337.916-00.477 (AGAL337.9-S), located 3.48\,kpc from the Sun.
The outflow lobes have extensions of less than 1\,pc---and thus were not fully resolved in the angular resolutions of ASTE and Mopra---and masses of 35\,--\,40\,$M_\odot$.
The maximum velocities of the outflow lobes are as high as 35\,--\,40\,$\kms$.
Our analysis of the infrared and sub-mm data indicates that AGAL337.9-S is in an early evolutionary stage of the high-mass star formation, having the total far-infrared luminosity of $\sim5\times10^4\,L_\odot$.
We also found that another dust core AGAL\,G337.922-00.456 (AGAL337.9-N) located 2$'$ north of AGAL337.9-S is a high-mass young stellar object in an earlier evolutional stage than AGAL337.9-S, although it is less bright in the mid-infrared than AGAL337.9-S.

\end{abstract}





\section{Introduction} \label{sec:intro}
High-mass stars play an important role in galactic evolution by releasing a significant amount of energy into their natal clouds and the local interstellar medium.
Although there have been a number of theoretical studies discussing how it is possible to form high-mass stars up to 140\,$M_\odot$ through disk-mediated accretion \citep{kru2009, hos2009, kui2010,kui2011}, there is a lack of observational study probing the evolutionary stages of mass accretion onto high-mass protostars.

\begin{figure*}[ht!]
\figurenum{1}
\epsscale{1.0}
\plotone{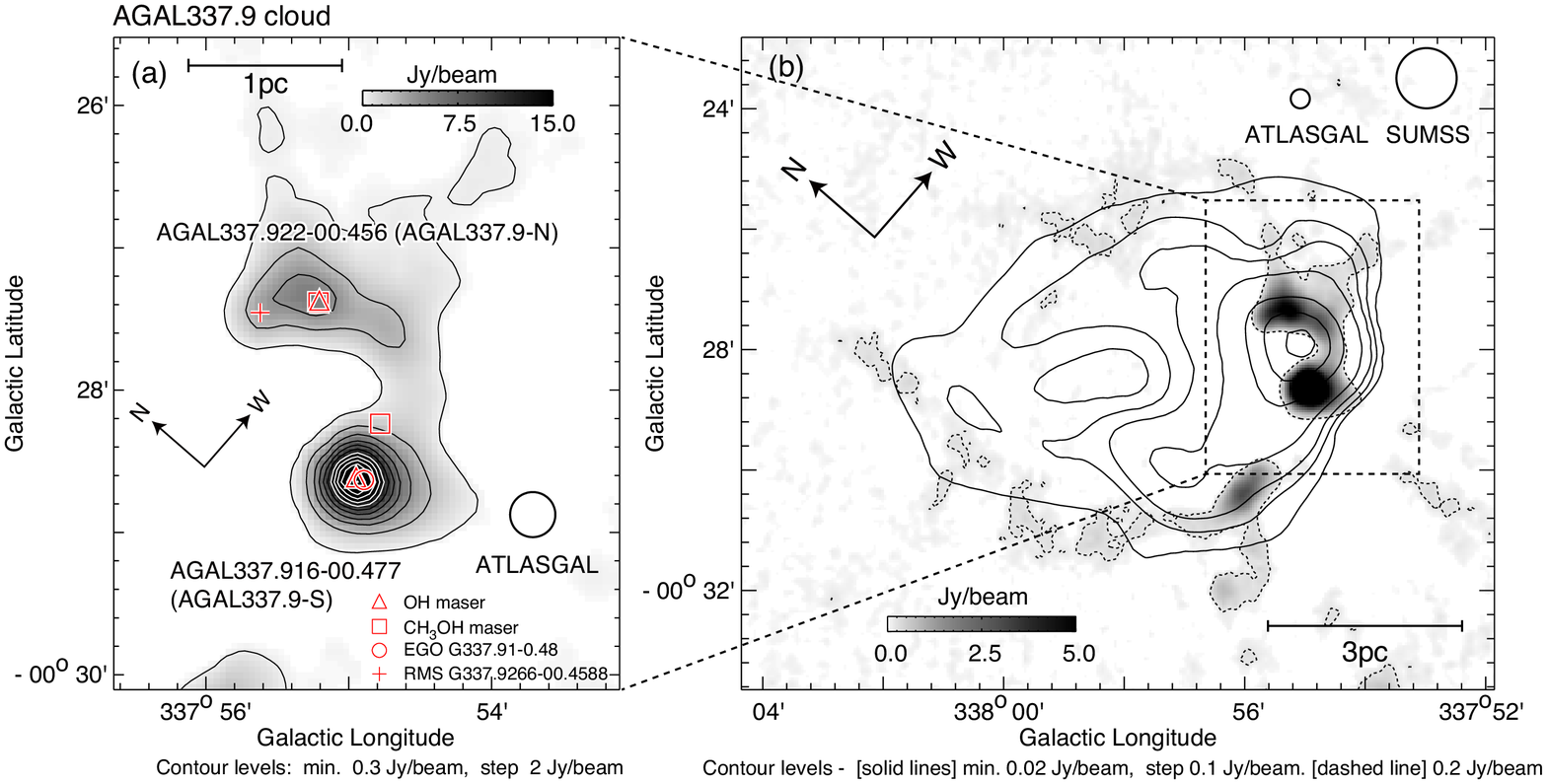}
\caption{(a) Distribution of the ATLASGAL 870\,$\mu$m emission \citep{urq2014} near AGAL337.9-S and AGAL337.9-N. EGO\,G337.91-0.48 is indicated by a cross \citep{cyg2008}. OH and CH$_3$OH masers are depicted with triangles and squares, respectively \citep{cas1998,cas2011,val2000}, and RMS\,G337.9266-00.4588, which indicates a massive YSO candidate, is shown as a cross \citep{mot2007}. EGO\,G337.91-0.48 is shown with circle. (b) The {\it SUMSS} 848\,MHz radio continuum emission \citep{boc1999} is shown with solid line contours superimposed onto the large-scale 870\,$\mu$m image, which is outlined by dashed contours. The scale bars are plotted by assuming a distance of 3.48\,kpc. \label{fig:s36}}
\end{figure*}

An important signpost for possibly investigating mass accretion is bipolar molecular outflow (see reviews by \citealt{fuk1989}, \citealt{fuk1993}, \citealt{ric2000} and \citealt{arc2007} and references therein).
Statistical studies of such molecular outflows indicate that there are linear correlations between outflow energetics and source bolometric luminosity over two orders of magnitude ($\sim10^3$\,--\,$10^5\,L_\odot$), implying the existence of a common outflow-driving mechanism that scales with luminosity \citep[e.g.,][]{rod1982, bal1983, rid2001, beu2002, mau2015a}. However, as these studies involved only a handful of samples with high luminosities on the order of $\sim10^5\,L_\odot$, it will be necessary to enlarge the sample of massive molecular outflows, particularly those erupting from young O-type stars which have not settled in mass growth, to understand the evolutionary scenario of high-mass star formation through a characterization of the infalling envelope, jet, outflow, or even the central object.

In this study, we report the discovery of new massive outflow associated with the dense dust core AGAL337.916-00.477.
AGAL337.916-00.477 was identified in the 870\,$\mu$m dust continuum survey, APEX Telescope Large Area Survey of the Galaxy (ATLASGAL) \citep{urq2014, con2013} (Figure\,\ref{fig:s36}(a)). 
AGAL337.916-00.477 is linked with another ATLASGAL source to the north, AGAL337.922-00.456, by diffuse 870\,$\mu$m emission with a curved shape.
For the sake of convenience, we hereafter refer to AGAL337.916-00.477 and AGAL337.922-00.456 as ``AGAL\,337.9-S'' and ``AGAL\,337.9-N,'' respectively, and refer to the entirety of the 870\,$\mu$m structure including AGAL337.9-S, AGAL337.9-N, and the diffuse emission connecting these as ``the AGAL337.9 cloud'.

The AGAL337.9 cloud shows several signs of high-mass star formation (see Figure\,\ref{fig:s36}(a)).
The extended green object EGO\,G337.91-0.48 identified by \citet{cyg2008}, which indicates a massive molecular outflow candidate, is found toward the peak of AGAL337.9-S. 
OH and CH$_3$OH maser sources are found toward the peak positions of both AGAL337.9-S and AGAL337.9-N with estimated positional errors of 0.4$''$ \citep{cas1998,cas2011} , and a young stellar object (YSO) candidate Red MSX Source (RMS) G337.9266-00.4588 has been identified to the east of AGAL337.9-N with a positional error of 2$''$ \citep{lum2002, mot2007}.
In addition, a cluster candidate, [DBS2003]\,172, was identified near the AGAL337.9 cloud by \citet{dut2003}, although $H$ and $K_{\rm S}$ photometric observations by \citet{bor2006} indicate that it is not a cluster but an extended star forming region without any O stars.

The kinematic distance of AGAL337.9-S and AGAL337.9-S was measured by \citet{wie2015}. The authors divided ATLASGAL sources into multiple peaks and determined a distance by measuring the radial velocities of those peaks with molecular line observations. The distance was then estimated by assuming the Galactic rotation curve by \citet{bra1993}. 
The authors adopted the near-side kinematic distance of $3.48\pm0.36$\,kpc to the complex including AGAL337.9-S and AGAL337.9-N by analyzing the H{\sc i} absorption. Following their study, in this paper we apply a distance 3.48\,kpc to the AGAL337.9 cloud.

The AGAL337.9 cloud is embedded in the south-west of the H{\sc ii} region, G337.90-00.50. 
The contour map with solid lines in Figure\,\ref{fig:s36}(b) shows the 848\,MHz radio continuum emission obtained by \citet{boc1999} superimposed on the 870\,$\mu$m image.
G337.90-00.50 has a size of $8' \times 6'$ and is surrounded by the 870\,$\mu$m component, which form a ring-like structure.
\citet{hat2016} measured the total infrared luminosity of G337.90-00.50 to be $1.7\times10^6$\,$L_\odot$ at 3.48\,kpc, where AGAL337.9-S was masked owing to errors in making fits for the Spectral Energy Distributions (SEDs), although the exciting source of the ionized gas in G337.90-00.50 has not to date been identified.

\begin{table*}
\begin{center}
\caption{Parameters of infrared/sub-mm data} \label{tab:1}
\begin{tabular}{ccccc}
\hline
\hline
Satellite/Telescope & Detector/Receiver & Band ($\mu$m)& Resolution ($''$) & Reference\\
(1) & (2) & (3)  & (4)\\
\hline
\multirow{4}{*}{{\it Spitzer}}	& \multirow{4}{*}{\it IRAC}	& 3.6		& 1.7	 	& \multirow{4}{*}{[1, 2]}\\
						&					& 4.5		& 1.7 	& \\
						&					& 5.8		& 1.9 	&\\
						&					& 8.0		& 2.0 \\
\hline
{\it AKARI}					& {\it IRC}				& 18		& 4 		& [3, 4] \\
\hline
\multirow{5}{*}{{\it Herschel}}	& \multirow{2}{*}{\it PACS}	& 70		& 5.2 	& \multirow{5}{*}{[5, 6, 7, 8]}\\
						&					& 160	& 12 \\
						\cline{2-4}
						&  \multirow{3}{*}{\it SPIRE}	& 250	& 18 \\
						&					& 350	& 25	\\
						&					& 500	& 37 \\
\hline					
{\it APEX}					& {\it LABOCA}			& 870	& 19.2	& [9] \\

\hline
\end{tabular}
\tablecomments{References [1] \citet{ben2003}, [2] \citet{chu2009}, [3] \citet{hat2016} [4] Ishihara et al. (2017) in preparation. [5] \citet{mol2010}, [6] \citet{pil2010}, [7] \citet{gri2010}, [8] \citet{pog2010}, and [9] \citet{urq2014}}
\end{center}
\end{table*}

During 2014 we made new CO $J$=3--2 and $J$=1--0 observations of the AGAL337.9 cloud using the Atacama Submillimeter Telescope Experiment (ASTE) and Mopra telescopes at beam sizes of $22''$\,--$35''$. 
In Section \ref{sec:obs} we describe the observations and dataset used in this study, and in Section \ref{sec:res} we present the main results of our new CO observations and the corresponding distributions of infrared emission. We also derive the physical properties of the molecular outflow and AGAL sources.
In Section \ref{sec:dis} we discuss the results and as summary is presented in Section \ref{sec:con}.

\section{Datasets} \label{sec:obs}
\subsection{ASTE CO $J$=3--2 observations}  \label{sec:obs:1}
Observations in the $^{12}$CO $J$=3--2 transition were performed on a $4'\times4'$ area centered on the AGAL337.9 cloud using the ASTE 10-m telescope located in Chile on the 28th and 29th of August 2014 \citep{eza2004, eza2008, ino2008}. 
The waveguide-type sideband-separating SIS mixer receiver ``CATS345'' and digital spectrometer ``MAC'' were used in the narrow-band mode, which provides 128\,MHz bandwidth and 0.125\,MHz resolution \citep{sor2000}. 
The corresponding velocity coverage and velocity separation at 345\,GHz are 111$\kms$ and 0.11$\kms$, respectively. 
The beam size was 22$''$ at 345\,GHz, and the observations were made in the on-the-fly (OTF) mode at a grid spacing of 7.5$''$.
The pointing accuracy was checked approximately every 1.5\,h to maintain pointing within 5$''$ by observing RAFGL\,4211 (R.A., Dec.)=($15^{\rm h}11^{\rm m}41\fs44$, $-48\degr19\arcmin59\farcs0$).
Absolute intensity calibration was performed using observations of W28 (R.A., decl.)=($17^{\rm h}57^{\rm m}26\fs8$, $-24\degr03\arcmin54\farcs0$). The day-to-day fluctuations of the peak intensity were within 10\,\%.
The typical system temperature was $\sim$250\,K in the single sideband (SSB) and the final root-mean-square (r.m.s) noise fluctuations were typically 0.2\,K at an output velocity resolution 0.44$\kms$.

\subsection{Mopra CO $J$=1--0 observations}  \label{sec:obs:2}
Observations in the $^{12}$CO $J$=1--0, $^{13}$CO $J$=1--0 and C$^{18}$O $J$=1--0 transitions were made using the Mopra 22-m telescope in Australia on the 25th of July 2014.
The velocity coverage and resolution provided by the ``MOPS'' backend system were $360\kms$ and $0.08\kms$, respectively, at 115\,GHz.
The OTF mode was used to cover the $4'\times4'$ area of the AGAL337.9 cloud, and the obtained spectra were spatially smoothed to a 45$''$ beam size and gridded to a 15$''$ spacing. The final velocity resolution was 0.44$\kms$.
The pointing accuracy was kept within 7$''$ by observing 86\,GHz SiO masers approximately every 1\,h.
Absolute intensity calibrations were performed by comparing the CO spectra calibrated using an ``extended'' beam efficiency \citep{lad2005} with observations of Orion-KL (R.A., Dec.)=($-5^{\rm h}35^{\rm m}14\fs5$, $-5\degr22\arcmin29\farcs6$).
The typical r.m.s. noise fluctuations of the $^{12}$CO, $^{13}$CO and C$^{18}$O $J$=1--0 emissions were 0.6\,K, 0.4\,K and 0.4\,K, respectively, with a typical system noise temperature of 400\,--\,600 K in the SSB.

\begin{figure*}
\figurenum{2}
\epsscale{.8}
\plotone{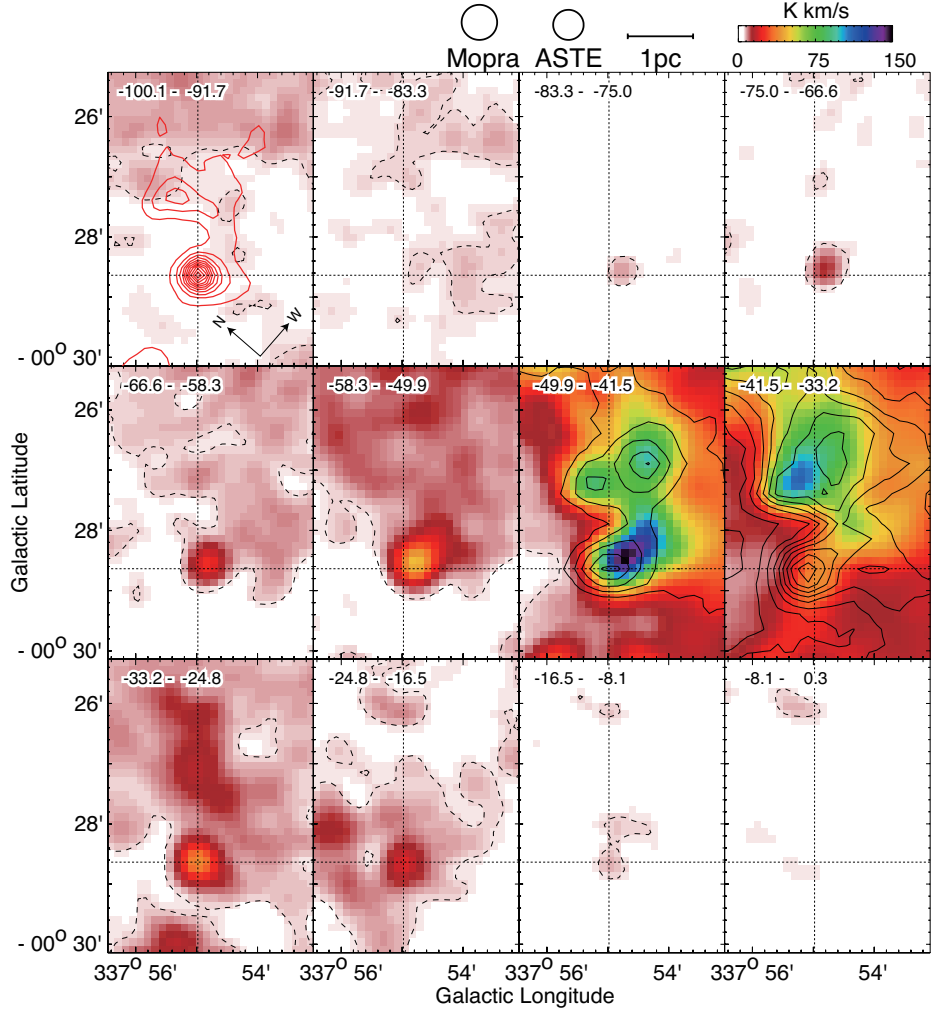}
\caption{Velocity channel maps of the $^{12}$CO $J$=3--2 emissions. The velocity range in each panel is shown at the left-top of the panel in units of km\,s$^{-1}$. The background image is the $^{12}$CO $J$=3--2 transition, in which the 3\,$\sigma$ level (2.5\,K\,km\,s$^{-1}$) is outlined by dashed contours.
The solid contours show the $^{13}$CO $J$=1--0 transition and are plotted every 6\,K\,km\,s$^{-1}$ from 10\,K\,km\,s$^{-1}$.  For comparison, the contour map of the ATLASGAL 870\,$\mu$m emission is plotted in red contours at the left-top of the panel, and the peak position of AGAL337.9-S in the 870\,$\mu$m distributions is indicated by two orthogonal dotted lines in each panel. \label{fig:ch}}
\end{figure*}

\subsection{Archival dataset}  \label{sec:obs:3}
We utilized the infrared and sub-mm datasets covering a wide range of wavelengths from 3.6\,$\mu$m to 870\,$\mu$m.
The near-infrared wavelengths were covered by the {\it Spitzer/IRAC} GLIMPSE survey results for the 3.6, 4.5, 5.8, and 8.0\,$\mu$m bands \citep{ben2003, chu2009}. 
The mid-infrared 18\,$\mu$m image was obtained from the {\it AKARI/IRC} all-sky survey dataset (\citealt{hat2016}; Ishihara et al. 2017 in preparation). 
Note that the {\it Spitzer/MIPS} MIPSGAL 24\,$\mu$m image \citep{car2009} is saturated in a large area toward the AGAL337.9-S cloud and therefore was not used in the present study.
The Hi-GAL survey carried out using the {\it PACS} and {\it SPIRE} instruments equipped on {\it Herschel} provides far-infrared maps at 70, 170, 250, 350, and 500\,$\mu$m \citep{mol2010, pil2010, gri2010,pog2010}.
As already shown in Figure\,\ref{fig:s36}(a), the sub-mm 870\,$\mu$m map was taken from the {\it APEX/LABOCA} ATLASGAL survey \citep{urq2014}.
The parameters of these infrared and sub-mm data are summarized in Table\,\ref{tab:1}.
In addition to these infrared and sub-mm maps, the {\it SUMSS} 848\,MHz radio continuum data \citep{boc1999} obtained at a beam size of $\sim1'$ was used to probe the ionized gas in G337.90-00.50.

\begin{figure*}
\figurenum{3}
\epsscale{1.}
\plotone{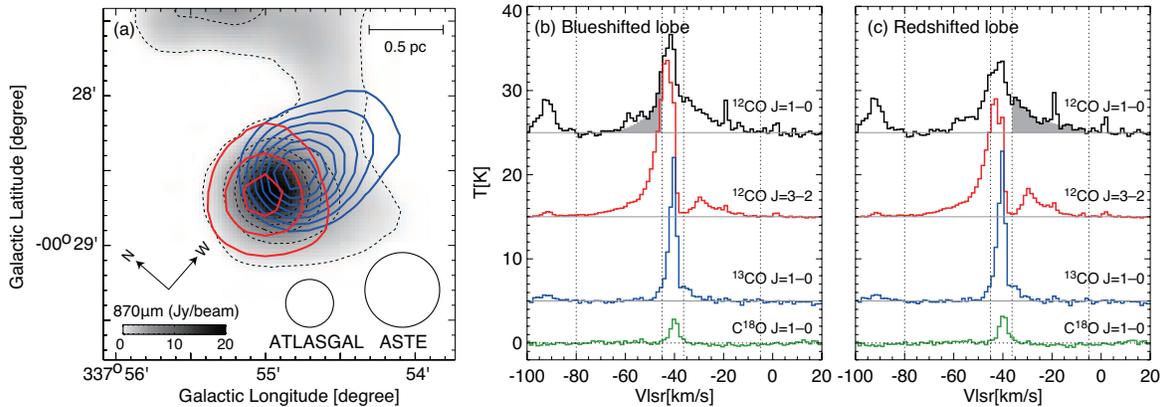}
\caption{ (a) $^{12}$CO $J$=3--2 distributions of the molecular outflow in AGAL337.9-S, where the blueshifted lobe is shown in blue contours and the redshifted lobe is in red contours. Background image and dashed contours are the ATLASGAL 870\,$\mu$m emissions, with contours plotted every 3\,Jy\,beam$^{-1}$ from 0.3\,Jy\,beam$^{-1}$. (b, c) The averaged CO spectra of the redshifted and blueshifted lobes, respectively, with the spectra smoothed to  represent a uniform velocity resolution of 1.3$\kms$. The shaded area in the $^{12}$CO $J$=1--0 was used for the mass estimates (see Appendix A). \label{fig:co}}
\end{figure*}

\section{Results} \label{sec:res}
\subsection{Molecular outflow in AGAL337.9-S} \label{sec:res:1}
Using the CO dataset obtained from ASTE and Mopra, we discovered a bipolar molecular outflow toward AGAL337.9-S.
Figure\,\ref{fig:ch} shows the velocity channel maps of the $^{12}$CO $J$=3--2 and $^{13}$CO $J$=1--0 emissions at a velocity interval of $\sim8.4\kms$.
The molecular gas associated with the AGAL337.9 cloud is pronounced at a velocity range of $\sim-50$\,--\,$-30$\,km\,s$^{-1}$, with the blue- and red-shifted lobes of the molecular outflow seen at $\sim-80$\,--\,$-45\kms$ and $\sim-35$\,--\,$-5\kms$, respectively.  The full width at zero intensity (FWZI) is thus measured as $\sim75\kms$.

Figure\,\ref{fig:co}(a) shows the $^{12}$CO $J$=3--2 contour maps of the blueshifted (blue contours) and redshifted (red contours) lobes superimposed on the 870\,$\mu$m image (Details of the lobe identifications are discussed in Appendix A).
The blueshifted lobe is slightly elongated to the west of the 870\,$\mu$m peak of AGAL337.9-S, while the redshifted lobe almost exactly coincides with the 870\,$\mu$m peak. 
The two lobes are barely resolved at the present resolution of the ASTE observations, and we therefore cannot discern the collimation of the outflow.
The CO spectra averaged over the outflow lobes are shown in Figures\,\ref{fig:co}(b) and (c), respectively, in which very broad wing features are clearly seen in the $^{12}$CO $J$=1--0 and $J$=3--2 emissions. 
On the other hand, there is no clear evidence of the molecular outflow in the other parts of the AGAL337.9 cloud, including AGAL337.9-N.

\begin{table*}
\begin{center}
\caption{Physical parameters of the observed molecular outflow.} \label{tab:2}
\begin{tabular}{lccc}
\hline
\hline
Parameter & Blueshifted lobe & Redshifted lobe &  Total/Average ($^{\rm t}/^{\rm a}$)\\
(1) & (2) & (3)  & (4)\\
\hline
maximum velocity, $v_{\rm max}$ (km\,s$^{-1}$) & 39.5 & 35.5 & 37.5\tablenotemark{a}\\
maximum radius, $r_{\rm max}$ (pc) & 0.83 & 0.41 & 0.62\tablenotemark{a}\\
dynamical time, $t_{\rm dyn}$ ($10^4$\,yr) & $<2.0$ & $<1.1$ & ---\\
Outflow mass, $M_{\rm flow}$ ($M_\odot$) & 53.7 & 54.3 & 108.0\tablenotemark{t} \\
Outflow momentum, $P_{\rm flow}$ ($M_\odot$\,km\,s$^{-1}$) & 568.6 & 771.3 & 1339.9\tablenotemark{t}\\
Outflow energy, $E_{\rm flow}$ ($\times10^{46}$\,ergs) & 7.4 & 14.0 & 21.5\tablenotemark{t} \\
\hline
\end{tabular}
\tablecomments{$^t$ and $^a$ stand for the total and the average values, respectively, of the parameters measured for the blueshifted and redshifted lobes, respectively.}
\end{center}
\end{table*}

\subsection{Physical parameters of the molecular outflow in AGAL337.9-S} \label{sec:res:2}
Here we estimate the physical parameters of the molecular outflow in AGAL337.9-S.  
Details of the estimates are described in Appendix A and are listed in Table\,\ref{tab:2}; in this subsection, we briefly summarize the results.

The systemic velocity $v_{\rm sys}$ of AGAL337.9-S was measured to be $\sim-40.5\kms$ using the optically thin C$^{18}$O data.
The velocity ranges of the two outflow lobes were determined to be $\sim-80$\,--\,$-45.1\kms$ and $\sim-36.3$\,--\,$-5\kms$ for the blueshifted and redshifted lobes, respectively. 
The corresponding maximum velocities $v_{\rm max}$ of the two lobes were then measured to be $|-80 - v_{\rm sys}| = 39.5\kms$ and $|v_{\rm sys} - -5| = 35.5\kms$, respectively.

The derived masses of the outflow lobes $M_{\rm flow}$ were 53.7\,$M_\odot$ and 54.3\,$M_\odot$ for the bleushfited and redshifted lobes, respectively, nearly the same for both.
These derived masses were then used to estimate the momentum $P_{\rm flow}$ and energy $E_{\rm flow}$ as 568.6\,$M_\odot\kms$ and $7.4\times10^{46}$\,ergs for the blueshifted lobe and 771.3\,$M_\odot\kms$ and $14.0\times10^{46}$\,ergs for the redshifted lobe.

The dynamical timescales $t_{\rm dyn}$ of the two outflow lobes were estimated using $v_{\rm max}$ and the physical lengths of the outflow lobes.
Because neither lobe was sufficiently resolved in the present dataset (see Figure\,\ref{fig:co}(a)), it was difficult to measure their lengths accurately.
It was also difficult to determine the angle of the outflow axis relative to the line-of-sight in order to obtain the ``real'' physical lengths from the ``projected'' physical lengths. 
We therefore give only lower limits of $t_{\rm dyn}$ as $2.0\times10^4$\,yrs and $1.1\times10^4$\,yrs for the blueshifted and redshifted lobes, respectively.

\subsection{The AGAL337.9 cloud in the infrared and sub-mm maps} \label{sec:res:3}
\begin{figure*}
\figurenum{4}
\epsscale{.6}
\plotone{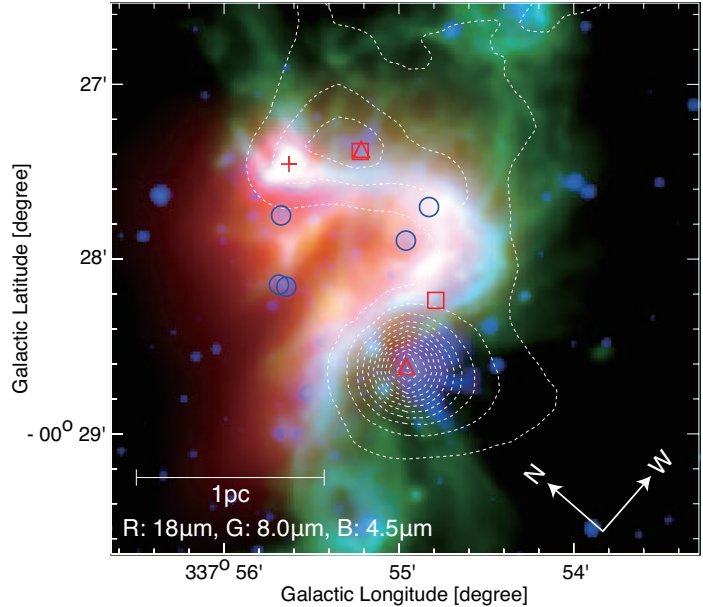}
\caption{Three color composite images of the infrared emissions in the AGAL337.9 cloud. Red: {\it AKARI/IRC} 18\,$\mu$m, green: {\it Spitzer/IRAC} 8\,$\mu$m, and blue: {\it Spitzer/IRAC} 4.5\,$\mu$m. A contour map of the ATLASGAL 870\,$\mu$m emission is superimposed, with the contours with dashed lines plotted every 3\,Jy\,beam$^{-1}$ from 0.3\,Jy\,beam$^{-1}$. The plotted symbols in red lines are the same as in Figure\,\ref{fig:s36}(a) but EGO\,337.91-0.48, while the blue circles indicate the positions of the candidate exciting sources of G337.90-00.50 (see Section\,\ref{sec:dis:4}).\label{fig:ir}}
\end{figure*}

Figure\,\ref{fig:ir} shows a composite image of the 18 (red), 8.0 (green), and 4.5\,$\mu$m (blue) data in the AGAL337.9 cloud.
As listed in Table\,\ref{tab:1} the spatial resolutions of those three bands are much higher than that of the 870\,$\mu$m data.
The 8.0\,$\mu$m emission is pronounced at the eastern rim of the cloud, which can be attributed to polycyclic aromatic hydrocarbon (PAH).
PAHs are excited by ultra-violet (UV) radiation from high-mass stars.
The 18\,$\mu$m emission is extended at the inside of the bent distributions of the AGAL337.9 cloud, which indicates warm dust grains being heated by the H{\sc ii} region G337.90-00.50.
These characteristics suggest interaction between the H{\sc ii} region and the AGAL337.9 cloud.
Erosion by the H{\sc ii} region seems to have not reached the central parts of AGAL337.9-S and AGAL337.9-N, as there is no significant enhancement of the 8 and 18\,$\mu$m emissions in these regions.
The 4.5\,$\mu$m emission, however, shows bright emission radially extended toward the south-west from the peak of AGAL337.9-S, which is cataloged as EGO\,337.91-0.48 in \citet{cyg2008}.
Because the {\it Spitzer/IRAC} 4.5\,$\mu$m band contains both H$_2$ ($\nu$=0--0, S(9, 10, 11)) lines and CO $\nu$=1--0 band heads which can be excited in the shocked molecular gas, the EGO sources are thought to be massive outflow candidates.
Spatial coincidence between EGO337.91-0.48 and AGAL337.9-S lends more credence to the existence of a massive molecular outflow, although the present resolution of the ASTE CO $J$=3--2 observations is not high enough to discuss physical interaction between the outflow lobes and EGO337.91-0.48.
A compact bright source is seen to the west of AGAL337.9-N, corresponding to the massive YSO candidate RMS\,G337.9266-00.4588 \citep{mot2007}, whereas it is not clear in the 870\,$\mu$m image.

\begin{figure*}
\figurenum{5}
\epsscale{.6}
\plotone{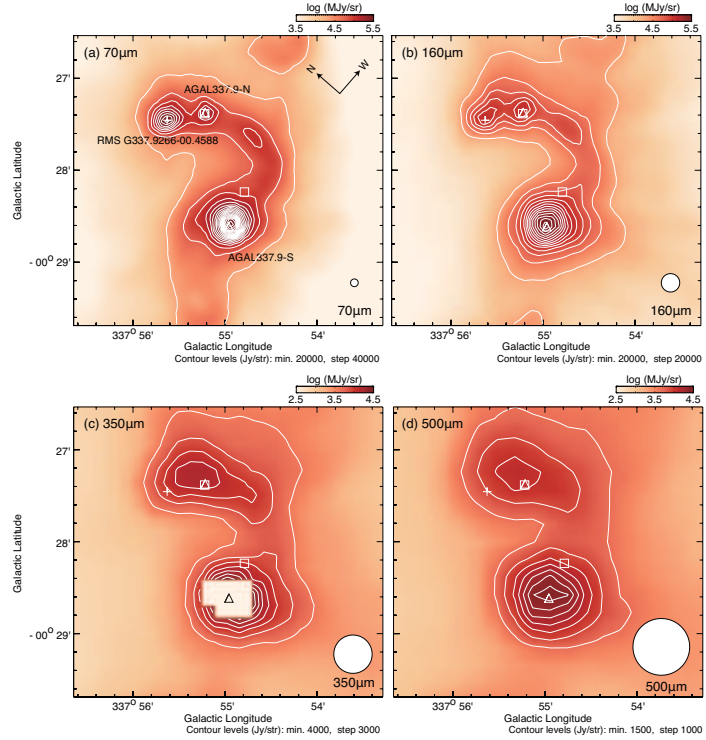}
\caption{Distributions of the {\it Herschel} (a) 70, (b) 160, (c) 350, and (d) 500\,$\mu$m emission in the AGAL337.9 cloud. The plotted symbols in (a) are the same as in Figure\,\ref{fig:s36}(a). \label{fig:ir2}}
\end{figure*}

Figure\,\ref{fig:ir2} shows the {\it Herschel} 70, 160, 350, and 500\,$\mu$m images. 
The four bands show distributions essentially similar to the 870\,$\mu$m emission shown in Figure\,\ref{fig:s36}(a), revealing the bent shape of the AGAL337.9 cloud and bright emission in AGAL337.9-S, where AGAL337.9-S is saturated in the 350\,$\mu$m image.
(note that both AGAL337.9-S and AGAL337.9-N are saturated in the 250\,$\mu$m image, which has therefore been removed from the analyses in this study). 
In the 70 and 160\,$\mu$m maps, RMS\,G337.9266-00.4588 shows bright and compact emission and can be distinguished from AGAL337.9-N, while it is ambiguous in the 350 and 500\,$\mu$m maps as well as in the 870\,$\mu$m map.

We estimated the angular sizes of AGAL337.9-S, AGAL337.9-N, and RMS\,G337.9266-00.4588 using the far-infrared and sub-mm maps by measuring the 1$\sigma$ radii deconvolved with the beam sizes (Table\,\ref{tab:1}).  
The size of AGAL337.9-S ranges from 6$''$--9$''$ ($\sim$ 0.10--0.15\,pc) in the 870, 160, and 70\,$\mu$m maps (a 3$''$ difference is within the pixel sizes of the images).
AGAL337.9-S and RMS\,G337.9266-00.4588 show circular shapes in the 160 and 70\,$\mu$m maps, with sizes of 5$''$--7$''$ and 6$''$--8$''$, respectively, while in the longer wavelength images only one component, which has slightly elongated shape, is seen (Figure\,\ref{fig:ir2}(c)).
The separation between AGAL337.9-S and RMS\,G337.9266-00.4588 measured in the 160 and 70\,$\mu$m maps is about 30$''$ ($\sim$ 0.5\,pc), which is not large enough to resolve at the beam sizes of the 350\,--\,870\,$\mu$m maps.
However, the peak position in these images is closer to AGAL337.9-N than RMS\,G337.9266-00.4588, and contamination from RMS\,G337.9266-00.4588 to the total flux in the northern part of the AGAL337.9 cloud may become relatively minor in the longer wavelengths.

\begin{figure*}
\figurenum{6}
\epsscale{.75}
\plotone{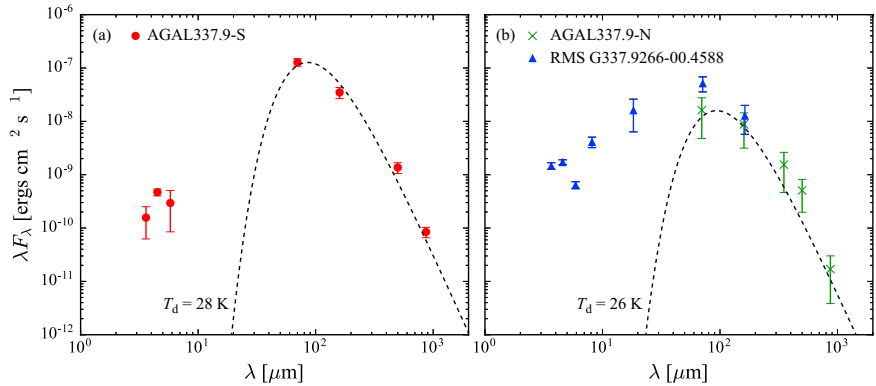}
\caption{SEDs obtained for (a) AGAL337.9-S and (b) AGAL337.9-N.
The best-fit curves of the modified-blackbody fits are shown as dashed lines. \label{fig:sed}}
\end{figure*}

\subsection{Spectral Energy Distributions} \label{sec:res:4}
We constructed the SEDs of the infrared sources embedded in the AGAL337.9 cloud using the available infrared and sub-mm maps.
For AGAL337.9-S, the 3.6, 4.5, 5.8, 70, 160, 500, and 870\,$\mu$m images were used, with circular apertures of radius 6$''$, 6$''$, 6$''$, 13$''$, 16$''$, 34$''$, and 20$''$ applied to the respective images. 
The aperture radii were determined by calculating the 2$\sigma$ width of AGAL337.9-S measured in the previous subsection convolved with the beam sizes.
In this analysis, the 8 and 18\,$\mu$m maps were removed, because they are too weak in AGAL337.9-S in comparison with its surroundings.
In the photometries, the local background level was estimated around the source using a circular annulus with an inner radius set to the same size as that of the circular aperture and an outer radius 1.5 times larger than the inner radius.
The median value of the pixels within the annulus was applied to the local background level, with the standard deviation used as the uncertainty of the photometry.
The results are plotted in Figure\,\ref{fig:sed}(a).
The SED is primarily dominated by far-infrared emission with a relative minor near- to mid-infrared emission.
We made a fit for the SED data points in the 70\,--\,870$\mu$m range using the modified-blackbody function with a fixed $\beta = 2$. The best-fit curve is shown in Figure\,\ref{fig:sed}(a) as a dashed line.
The dust temperature $T_{\rm dust}$ was derived as 28\,K, and the total far-infrared luminosity $L_{\rm FIR}$ was as high as $5.0\times10^4$\,$L_\odot$ at 3.48\,kpc.

The SED for AGAL337.9-N was built using 70, 160, 350, 500, and 870\,$\mu$m images.
The radii of the circular apertures were set to 11$''$, 14$''$, 23$''$, 33$''$, and 20$''$ respectively in these five bands. 
The local background was subtracted in the same manner as was done for AGAL337.9-S.
The results and best-fit curve are shown in Figure\,\ref{fig:sed}(b) as crosses and a dashed line, respectively.
$T_{\rm dust}$ and $L_{\rm FIR}$ are derived to be 26\,K and $6.3\times10^3$\,$L_\odot$, respectively.
The data points at 350 and 500\,$\mu$m were not included in the fit, because contamination by RMS\,G337.9266-00.4588 may be relatively large compared in these two bands with the 870\,$\mu$m band, although there may be contamination even in the latter.

For comparison, the SED for RMS\,G337.9266-00.4588 was constructed by performing aperture photometries in the 8.0, 18, 70, and 160\,$\mu$m images (Figure\,\ref{fig:ir2}), using the same aperture sizes as were used for AGAL337.9-N.
In addition, the fluxes of the 3.6, 4.5, and 5.8\,$\mu$m bands were obtained from the GLIMPSE PSC catalog \citep{ben2003}.
The results are plotted in Figure\,\ref{fig:sed}(b) as triangles.
It is seen that the near-infrared emission is enhanced in RMS\,G337.9266-00.4588, suggesting that it is in an evolved stage of star formation relative to AGAL337.9-S and AGAL337.9-N.

\begin{table*}
\begin{center}
\caption{Physical parameters of AGAL337.9-S and AGAL337.9-N.} \label{tab:3}
\begin{tabular}{ccccc}
\hline
\hline
Data & \multicolumn2c{$N_{\rm core}$} & $M_{\rm core}$ & $n_{\rm core}$\\
\cline{2-3}
& average & peak \\
& ($\times10^{22}$\,cm$^{-2}$) & ($\times10^{22}$\,cm$^{-2}$) & $M_\odot$ & ($\times10^5$\,cm$^{-3}$) \\
(1) & (2) & (3)  & (4) & (5)\\
\hline
\multicolumn5c{\underline{AGAL337.916-00.477 (AGAL337.9-S)}}\\
870\,$\mu$m              &14.5 &36.5& 1040 & 1.3--4.5\\
C$^{18}$O $J$=1--0  &9.5 & 12.5 & 950 & 1.2--4.2\\
\multicolumn5c{\underline{AGAL337.922-00.456 (AGAL337.9-N)}}\\
870\,$\mu$m              &11.3 & 13.8 & 230 & 1.2\\
C$^{18}$O $J$=1--0  & 4.8 & 5.3 & 180  & 0.6\\

\hline
\end{tabular}
\end{center}
\end{table*}

\subsection{Physical parameters of AGAl337.9-S and AGAL337.9-N} \label{sec:res:5}

Using the $T_{\rm dust}$ derived in the fits for the SEDs, we estimated the H$_2$ column density $N_{\rm core}$ and the H$_{2}$ mass $M_{\rm core}$ of AGAL337.9-S from the 870\,$\mu$m image.
Details of the estimates are summarized in Appendix B and Table\,\ref{tab:3}. 
The derived $N_{\rm core}$ is as high as $\sim$3.7$\times10^{23}$\,cm$^{-2}$ at the peak, while $M_{\rm core}$ is 1040\,$M_\odot$.
The number density of the AGAL337.9-S $n_{\rm core}$ was derived to be 1.3\,--\,4.5\,$\times10^5$\,cm$^{-3}$ by assuming a sphere with a 2$\sigma$ radius of 0.2\,--\,0.3\,pc.

The $N_{\rm core}$ and $M_{\rm core}$ of AGAL337.9-S were also estimated from the Mopra C$^{18}$O data by assuming a local thermodynamic equilibrium (LTE).
The abundance ratio of [$^{16}$O]/[$^{18}$O] was calculated as $58.8 \times D_{\rm gc} + 37.1 \sim 370$ \citep{wil1994}, where $D_{\rm gc}$ is the galactocentric distance in kpc, leading to a derived $N_{\rm core}$ and $M_{\rm core}$ of  $1.3\times10^{23}$\,cm$^{-2}$ and $\sim$950\,$M_\odot$, respectively. The $n_{\rm core}$ was derived as 1.2\,--\,4.2\,$\times10^5$\,cm$^{-3}$. These estimated values are consistent with those derived from the 870\,$\mu$m image.

The estimates for the AGAL337.9-N were carried out in the same manner as done for AGAL337.9-S. The $M_{\rm core}$ ranges from 180\,$M_\odot$ for C$^{18}$O to 230\,$M_\odot$ for 870\,$\mu$m, while the $N_{\rm core}$ at the peak position ranges from 0.5\,--\,1.4$\times10^{23}$\,cm$^{-2}$. These figures are smaller than that in AGAL337.9-S.
The $n_{\rm core}$ was derived as 0.6\,--\,1.2\,$\times10^5$\,cm$^{-3}$ by assuming a radius of 0.2\,pc .

\section{Discussion} \label{sec:dis}
\subsection{Comparisons with other massive molecular outflows and massive cores} \label{sec:dis:1}

In the previous sections, we reported the discovery of a bipolar molecular outflow with a broad velocity width of 75$\kms$ in FWZI.
The driving source, AGAL337.9-S, has a total far-infrared luminosity of $0.5\times10^4\,L_{\odot}$.

Recently, \citet{mau2015b} and \citet{mau2015a} conducted statistical studies of massive YSOs and massive outflows based on the CO $J$=3--2 observations of a relatively unbiased mid-infrared bright sample covering total source luminosities $L_{\rm src}$ from $10^3\,L_\odot$ to $\sim$10$^5\,L_\odot$.
As shown in Figure\,5 in \citet{mau2015a}, the authors found linear correlations between the outflow parameters and $L_{\rm src}$, which can be scaled to those of the low-mass sources. 
Compared with their results, the AGAL337.9 outflow have properties typical to the previously known massive outflows in the Milky Way; the $M_{\rm flow}$ of the AGAL337.9 outflow can be plotted in the middle of the scatter of the sample distribution (see the left-top panel of Figure\,5 in \citealt{mau2015a}), while $P_{\rm flow}$ and $E_{\rm flow}$ are seen around the upper ends of the scatters in the same source luminosity, although they are still within the scatters (see the right-top and left-center panels).
\citet{mau2015b} also built a correlation between $M_{\rm core}$ and $M_{\rm flow}$ (see their Figure\,9), and the AGAL337.9-S outflow can be plotted within their scatter and appears to follow its linear scaling.

\subsection{AGAL337.9-S: a massive YSO} \label{sec:dis:2}
The observed properties presented in this study indicate that AGAL337.9-S is a massive YSO.
\citet{mol2008} studied the evolutionary track of high-mass stars using a core mass-source luminosity diagram.
According to their model calculations plotted in Figure\,9 of \citet{mol2008}, in which a star formation efficiency of 50\% is assumed, AGAL337.9-S would be in a transition stage between the actively accreting phase and the envelope clean-up phase.

In this study, we investigated the existence of the H{\sc ii} region in AGAL337.9-S in order to probe the evolutionary stage of the core.
In the SUMSS 848\,MHz map shown in Figure\,\ref{fig:s36}(b), in which the emission from the ionized gas in the H{\sc ii} region G337.90-00.50 is dominant, we find no clear enhancement of the emission toward AGAL337.9-S, giving only an upper-limit of 0.3\,Jy.
By assuming that a compact (C), an ultra-compact (UC), or a hyper-compact (HC) H{\sc ii} region is embedded in AGAL337.9-S, we can calculate the predicted 848\,MHz fluxes to be 4.7, 0.2, and 0.02\,Jy, respectively, where the sizes and the emission measures (EMs) of these types are taken from \citet{kur2005} as (size, EM) = (0.5\,pc, $10^7$\,pc\,cm$^{-6}$), (0.1\,pc, $10^7$\,pc\,cm$^{-6}$), and (0.03\,pc, $10^{10}$\,pc\,cm$^{-6}$) for the CH{\sc ii}, UCH{\sc ii}, and HCH{\sc ii} regions, respectively.
It is possible that a HCH{\sc ii} region, which could not be detected in the 848\,MHz map, exists in AGAL337.9-S.
The UCH{\sc ii} region is marginal, as its predicted flux is almost equal to the observed upper-limit. 
The possibility of a CH{\sc ii} region can be ruled out.
The weak emissions in the 8 and 18\,$\mu$m bands in Figure\,\ref{fig:ir} also support the present discussion, as these are excited through interaction with H{\sc ii} regions.

The above discussion concerning the 848\,MHz emission can also be applied to RMS\,G337.9266-00.4588 (Figure\,\ref{fig:ir2}), although it displays bright near-infrared emission (Figure\,\ref{fig:ir}), indicating that RMS\,G337.9266-00.4588 is in a more evolved stage than AGAL337.9-S.

\subsection{AGAL337.9-N: another massive YSO} \label{sec:dis:3}
The above observations and analyses indicate that AGAL337.9-N is a compact source with a high molecular column density (see Figure\,\ref{fig:ir2} and Table\,\ref{tab:3}) and a large infrared luminosity of $6.3\times10^3$\,$L_\odot$, suggesting that it is in an early stage of high-mass star formation.
This is consistent with the associations of OH and CH$_3$OH masers.
As seen in Figure\,\ref{fig:ir}, AGAL337.9-N has no significant emission in the near- and mid-infrared bands and is likely younger than AGAL337.9-S.
In the mass-luminosity diagram of \citet{mol2008}, the $M_{\rm core}$ and $L_{\rm tot}$ of AGAL337.9-S can be plotted below the turnover in the model evolutional tracks, lending more credence to this present assumption.

\subsection{H{\sc ii} region G337.90-00.50} \label{sec:dis:4}
\begin{figure*}
\figurenum{7}
\epsscale{.8}
\plotone{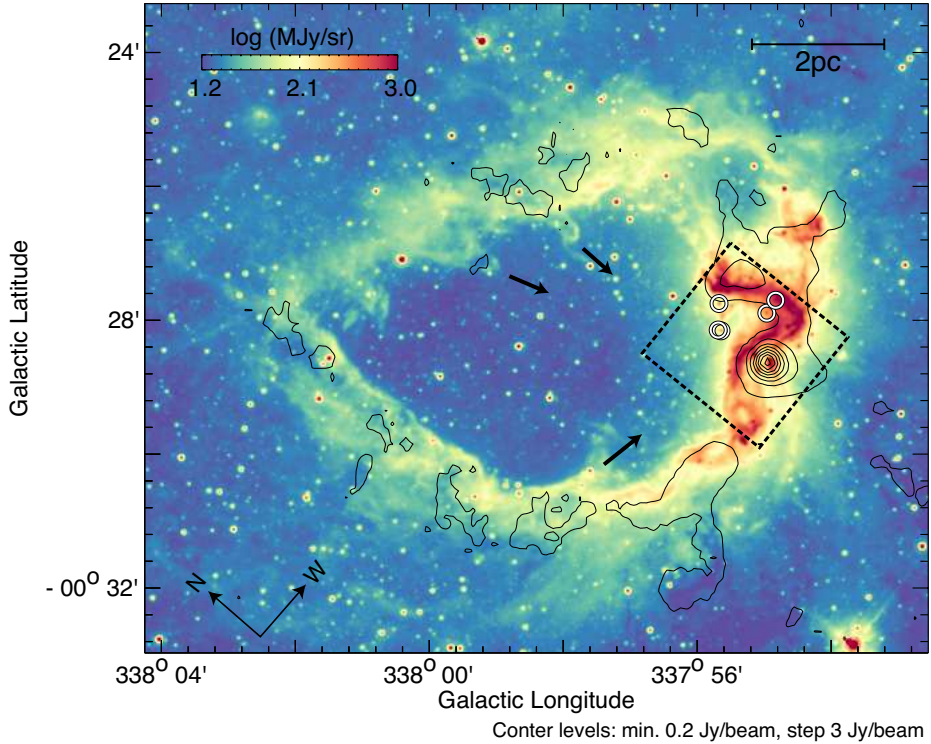}
\caption{{\it Spitzer/IRAC} 5.6\,$\mu$m image for a large area of G337.90-00.50 with the contour map of the 870\,$\mu$m distributions. The box with dashed lines shows the region observed by \citet{bor2006}. The arrows indicate the directions which the infrared pillars are pointing. The white circles indicate the candidate exciting sources of G337.90-00.50 (see text).\label{fig:irac8}}
\end{figure*}

As shown in Figure\,\ref{fig:s36}(b), the AGAL337.9 cloud is embedded around the rim of the H{\sc ii} region G337.90-00.50.
The total infrared luminosity of the H{\sc ii} region was derived by \citet{hat2016} as $1.7\times10^6$\,$L_\odot$ without AGAL337.9-S in which the fits for the SEDs failed.
This remarkably high luminosity is about one oder of magnitude larger than the estimates for AGAL337.9-S and AGAL337.9-N and corresponds to an O5-type star if a single object is assumed \citep{pan1973}.
Figure\,\ref{fig:irac8} shows a large-scale distribution of the 5.6 and 870\,$\mu$m emissions in G337.90-00.50.
The ring-like structure in the 5.6\,$\mu$m emission surrounding G337.90-00.50 has pillars pointing toward the AGAL337.9 cloud, as depicted by arrows in Figure\,\ref{fig:irac8}.
In addition, the 18\,$\mu$m and 848\,MHz emissions peak around the AGAL337.9 cloud as shown in Figures\,\ref{fig:s36}(b) and \ref{fig:ir}.
These observed signatures strongly suggest that the exciting source of G337.90-00.50 hides in the vicinity of the AGAL337.9 cloud.

$H$ and $K_{\rm S}$ photometric observations toward the cluster candidate [DBS2003]172, which coincides with the AGAL337.9 cloud, were carried out by \citet{bor2006}. Their field of view is depicted by dashed lines in Figure\,\ref{fig:irac8}.
The color-magnitude ($K_{\rm S}$ vs. $H-K_{\rm S}$) diagram in Figure\,16 of \citet{bor2006} indicate that many stars in this region have $H-K_{\rm S}$ of 1\,--\,2\,mag, which correspond to visual extinction $A_{\rm V}$ of 5\,--\,10\,mag.
The Mopra $^{12}$CO $J$=1--0 intensity around the 848\,MHz peak integrated over the all velocity components along the line-of-sight between the Sun and the AGAL337.9 cloud is derived as $\sim90$\,K\,km\,s$^{-1}$. The corresponding $A_{\rm V}$ can be estimated as $\sim10$\,mag by assuming a X(CO) conversion factor of $2\times10^{20}$\,cm$^{-2}$\,(K\,km\,s$^{-1}$)$^{-1}$ \citep{str1988} and the relations $N$(H+H$_2$) = $5.8\times10^{21} E(B-V)$\,cm$^{-2}$\,mag$^{-1}$ and $A_{\rm v} = 3.1 E(B-V)$ \citep{boh1978}. 
This is consistent with the results of \citet{bor2006}, and it is therefore possible that the majority of the stars identified by \citet{bor2006} except for the unreddened foreground stars are distributed at the same distance as the AGAL337.9 cloud. 
All the stars identified by \citet{bor2006} have $K_{\rm S}$ fainter than 12\,mag as shown in their color-magnitude diagram, indicating no signs of OB stars.
However, using the Two Micron All Sky Survey (2MASS) point source catalog \citep{skr2006}, we found five stars having $K_{\rm S}$ brighter than 10\,mag and $H-K_{\rm S}$ of $>0.5$\,mag around the peak of the 848\,MHz emission, which are depicted in Figure\,\ref{fig:ir} and \ref{fig:irac8} by circles. These stars coincide with the sources saturated in the $K_{\rm S}$-band image of \citet{bor2006} and are therefore not included in their color-magnitude diagram.
If any of these stars having large near-infrared fluxes are associated with G337.90-00.50, these would be primary candidates of the exciting source of G337.90-00.50. Further investigation based on new observations is necessary for better understanding. 

It is difficult to derive the evolutionary timescale of G337.90-00.50 without information on the exciting source (position and mass), but it is still possible to roughly estimate the timescale by calculating the crossing time of the ionized gas.
The diameter of G337.90-00.50 is measured to be $\sim$6\,pc in Figure\,\ref{fig:s36}(b), and if we assume the expanding velocity of the H{\sc ii} region corresponds to the sound speed of the ionized gas, $\sim10\kms$, the timescale can be estimated as $\sim$0.6\,Myr.

\subsection{Star formation in G337.90-00.50} \label{sec:dis:5}
In G337.90-00.50 there are at least two populations of high-mass stars: one is the exciting source of G337.90-00.50, which likely hides in the vicinity of the AGAL337.9 cloud: the other is the massive YSOs embedded in the AGAL337.9 cloud, i.e., AGAL337.9-S, AGAL337.9-N, and RMS\,G337.9266-00.4588.

A possible scenario to interpret this continuous high-mass star formation is the ``collect \& collapse'' model \citep{elm1977}, in which an expanding H{\sc ii} region sweeps up the surrounding neutral material and forms a dense molecular shell, which is followed by the fragmentations by the gravitational instability and formation of the next generation of stars. 
The total molecular mass of the AGAL337.9-S cloud can be estimated to be $\sim$3000\,$M_\odot$ at $T_{\rm dust}$ of 26\,K from the 870\,$\mu$m map.
By assuming a uniform distance of 1\,pc between the exciting source of G337.90-00.50 and the AGAL337.9 cloud, the initial gas density prior to the onset of the ionization can be calculated as $\sim10^5$\,cm$^{-3}$, where we calculated the volume of the AGAL337.9 cloud assuming a bent cylinder with a radius of 0.4\,pc and a length of 3\,pc.
In medium with such a high density, collect \& collapse may not work as the mechanism of star formation, as the medium is already dense enough to become gravitational unstable without further mass growth.
It is therefore more likely that the AGAL337.9 cloud was already formed before the birth of G337.90-00.50.
In this case, it can be interpreted that the bent shape of the AGAL337.9 cloud was formed through the expansion of the H{\sc ii} region of G337.90-00.50: AGAL337.9-S and AGAL337.9-N with high column density were shielded from the erosion, and less dense region between these where the 18\,$\mu$m emission is now excited was selectively dissipated (Figure\,\ref{fig:ir}).

Another possible scenario is the cloud-cloud collision model. 
Based on molecular observations, it is increasingly evident that cloud-cloud collision plays an important role in triggering high-mass star formation \citep{fur2009, oha2010, tor2011, tor2015, tor2017, fuk2014, fuk2015b, fuk2016, fuk2017a, fuk2017b, gon2017, dew2017}.
\citet{tor2015} showed that a collision between two dissimilar clouds can explain the ring-like molecular distribution surrounding an H{\sc ii} region.
Their CO observations in the galactic H{\sc ii} region RCW\,120 found no evidence of the expansion of the ring-like cloud, which would be expected under the collect \& collapse model.
In their scenario, such ring-like molecular structures can be understood as cavities created through cloud-cloud collision.
We argue that the present case is explicable as due to cloud-cloud collision similar to RCW\,120 \citep{tor2015}. 
The bubble like shape is common in the two objects; the high mass stars are near or on the bottom of the bubble, which is open to the other side of the location of high mass stars. A difference between RCW\,120 and the present case is in the number of high mass stars; i.e.,  one in RCW\,120 and three in the present case. 
This is possibly explained as due to the difference in the molecular column density of the initial cloud; the column density in RCW\,120 is 1\,--\,$3\times10^{22}$\,cm$^{-2}$, while that of the present case is $>10^{23}$\,cm$^{-2}$.
We will investigate the possibility of the cloud-cloud collision in G337.90-00.50 in a separate paper based on the CO dataset which covers the entire region of G337.90-00.50 as well as the present dataset (Torii et al. 2017, in preparation).

\section{Summary} \label{sec:con}
The conclusions of the present work are summarized as follows.

\begin{enumerate}
\item We carried out CO $J$=1--0 and $J$=3--2 observations of two ATLASGAL sources, AGAL337.916-00.477 (AGAL337.9-S) and AGAL337.922-00.456 (AGAL337.9-N) with Mopra and ASTE and discovered bipolar molecular outflow in AGAL337.9-S. The projected sizes of the outflow lobes are as less than 1\,pc and are not fully resolvable with the spatial resolutions of $\sim30''$ available with ASTE and Mopra.
\item We calculated the physical parameters of the molecular outflow, which are summarized in Table\,\ref{tab:2}.
The derived parameters are roughly consistent between the blueshifted and redshifted lobes of the outflow. In each lobe, the outflow mass $M_{\rm flow}$ is about 40\,$M_\odot$ and the maximum velocity $v_{\rm max}$ is as high as $\sim40\kms$.
\item We constructed a SED for AGAL337.9-S, as shown in Figure\,\ref{fig:sed}(a), and derived a total far-infrared luminosity $L_{\rm FIR}$ of $\sim5\times10^4\,L_\odot$. 
We estimated the molecular core mass $M_{\rm core}$ of AGAL337.9-S as $\sim1000\,M_\odot$ and a molecular column density $N_{\rm core}$ as high as 1\,--\,4$\times10^{23}$\,cm$^{-2}$.
\item Compared with massive outflows and massive cores measured in other studies, the corresponding characteristics of AGAL337.9-S are similar and generally follow the same linear scaling.
\item We investigated the infrared and sub-mm datasets in another source AGAL337.9-N, which is located to the north of AGAL337.9-S. Its SED shown in Figure\,\ref{fig:sed}(b) suggests that it is in an earlier evolutionary stage than AGAL337.9-N, as AGAL337.9-N has no significant emissions in the near- and mid-infrared maps. The derived $L_{\rm FIR}$ of AGAL337.9-N is $6.3\times10^3\,L_\odot$. 
\item AGAL337.9-S and AGAL337.9-N are both located at the western rim of the H{\sc ii} region, G337.90-00.50, and we discussed that the exciting source of G337.90-00.50 would be distributed in the vicinity of AGAL337.9-S and AGAL337.9-N.
Developing a through understanding of mechanisms of the continuous high-mass star formation and their corresponding histories is an important issue for the future study.
\end{enumerate}

\acknowledgments

The authors thank the anonymous referee for his/her helpful comments.  This work was financially supported by Grants-in-Aid for Scientific Research (KAKENHI) of the Japanese society for the Promotion of Science (JSPS; grant numbers 15H05694, 15K17607, 24224005, 26247026, 25287035, and 23540277). 
{\it Herschel} is an ESA space observatory with science instruments provided by European-led Principal Investigator consortia and with important participation from NASA. This work is based in part on observations made with the {\it Spitzer} Space Telescope, which is operated by the Jet Propulsion Laboratory, California Institute of Technology under a contract with NASA. This research is also based on observations with AKARI, a JAXA project with the participation of ESA. This publication makes use of data products from the Two Micron All Sky Survey, which is a joint project of the University of Massachusetts and the Infrared Processing and Analysis Center/California Institute of Technology, funded by the National Aeronautics and Space Administration and the National Science Foundation.

\appendix
\section{Estimates of the physical parameters of the molecular outflow in AGAL337.9-S}

As the CO spectra of the AGAL337.9-S outflow are mixtures of the outflow component and the core component, we first determined the velocity range of the core component.
The spatial extension of the core component was identified by drawing a contour on the $^{13}$CO $J$=1--0 integrated intensity map at two-third level of the peak intensity. 
A fit with a Gaussian function was performed for the $^{13}$CO $J$=1--0 spectrum averaged over the defined extension of the core component, providing a systemic velocity $v_{\rm sys}$ of $-40.5\kms$ and a velocity width $\sigma_{\rm v}$ of $1.5\kms$.
The velocity range of the core component was derived as $v_{\rm sys} \pm 3\,\sigma_{\rm v}$, with the velocity ranges of the blueshifted and redshifted lobes determined to be $-80$\,--\,$-45.1\kms$ and $-36.3$\,--\,$-5\kms$, respectively.
The maximum velocities $v_{\rm max}$ were calculated to be $|-80 - v_{\rm sys}| = 39.5\kms$ and $|v_{\rm sys} - -5| = 35.5\kms$ for the blueshifted lobe and the redshifted lobe, respectively (Table\,\ref{tab:1}).
The contour maps shown in Figure\,\ref{fig:co}(a) in red and blue are the $^{12}$CO $J$=3--2 distributions integrated over those two velocity ranges, where we defined the extension of each lobe at one-third level of the peak value. 
The averaged CO spectra of the lobes are shown in Figures\,\ref{fig:co}(b) and (c).

The mass estimates of the outflow lobes in the previous studies were carefully done by correcting optical depth effect of the CO emission. 
For example, \citet{mau2015a} took $^{12}$CO/$^{13}$CO intensity ratios of outflow lobes at every ($x, y, v$) voxels in a CO cube data and estimated their molecular masses.
The noise levels of the present $^{12}$CO $J$=1--0 and $^{13}$CO $J$=1--0 dataset are not as good as those in these studies. 
We therefore estimated the masses of the outflow lobes $M_{\rm flow}$ using the averaged $^{12}$CO $J$=1--0 spectra shown in Figures\ref{fig:co}(b) and (c) with an assumption of the uniform $^{12}$CO/$^{13}$CO ratio. 
We also assumed a single excitation temperature $T_{\rm flow, ex}$, which can be calculated using the optically thick $^{12}$CO $J$=3--2 emission from the equation
\begin{equation}
T_{\rm flow, ex} \ = \ \frac{16.59}{\ln [1 + 16.59 / (T_{3-2} + 0.038)]}, \label{eq:0a}
\end{equation}
where $T_{3-2} = 27$\,K was the observed peak temperature of the $^{12}$CO $J$=3--2 emission toward AGAL337.9-S. The derived $T_{\rm flow, ex}$ was 35\,K. 

The assumptions of the uniform $T_{\rm flow, ex}$ allowed us to calculate the optical depth of the outflow lobes using the following equation;
\begin{eqnarray}
\frac{T_{\rm flow, 12}}{T_{\rm flow, 13}} &=& \frac{1-\exp(-\tau_{\rm flow, 12})}{1-\exp(-\tau_{\rm flow, 13})} \\
 &=& \ \frac{1-\exp(-\tau_{\rm flow, 12})}{1-\exp{(-\tau_{\rm flow, 12}/R_{12/13})}}, \label{eq:1}
\end{eqnarray}
where $T_{\rm flow, 12}$ and $T_{\rm flow, 13}$ are the observed $^{12}$CO $J$=1--0 and $^{13}$CO $J$=1--0 intensities, and $\tau_{\rm flow, 12}$ and $\tau_{\rm flow, 13}$ are the optical depth of the $^{12}$CO $J$=1--0 and $^{13}$CO $J$=1--0 emissions, respectively. 
$R_{12/13}$ is the abundance ratio of [$^{12}$C]/[$^{13}$C] $\sim$ 50 derived from $R_{12/13} = 7.5 \times D_{\rm gc} + 37.1$ in the equation given by \citep{wil1994}, where $D_{\rm gc}$ is the distance from the Galactic center in kpc.
$T_{\rm flow, 12}$/$T_{\rm flow, 13}$ in the averaged CO $J$=1--0 spectra of the two lobes in Figures\,\ref{fig:co}(b) and (c) varies from 5 to 11, and we took an average of $T_{\rm flow, 12}/T_{\rm flow, 13} = 9$, with $\tau_{\rm flow, 12}$ estimated as 5.9 in equation\,\ref{eq:1}.

In the averaged $^{12}$CO $J$=1--0 spectra in Figures\,\ref{fig:co}(b) and (c), there are some narrow components overlapping the outflow components, i.e., the emissions around $-60\kms$ and $-53\kms$ in the blueshifted lobe and $-20\kms$ in the redshifted lobe.
These feaetures are not likely related to the outflow, and we here remove them from the averaged $^{12}$CO $J$=1--0 spectra by performing fits with a quadratic function. 
The resulting $^{12}$CO $J$=1--0 spectra are shown with shades in Figures\,\ref{fig:co}(b) and (c). 
By using these spectra, the molecular column density $N_{\rm flow}$ of the blueshifted lobe and the redshifted lobe at each velocity bin is calculated with the equation, 
\begin{equation}
 N_{\rm flow} = 4.3\times10^{13} \frac{T_{\rm flow, ex} + 0.92}{\exp(\frac{-5.53}{T_{\rm flow, ex}})} \frac{\tau_{\rm flow, 12}}{1-\exp({-\tau_{\rm flow, 12}})} T_{\rm flow, 12} dv, \label{eq:1a}
\end{equation}
where $dv$ is the channel resolution in $\kms$.
$M_{\rm flow}$ were then derived as
\begin{equation}
 M_{\rm flow} \ = \ \sum_i M_{{\rm flow}, i},
\end{equation}
where $i$ represents each velocity bin.
The derived $M_{\rm flow}$ were 53.7\,$M_\odot$ and 54.3\,$M_\odot$ for the bleushfited and redshifted lobes, respectively.

The momentum $P_{\rm flow}$ and the energy $E_{\rm flow}$ of the lobes were calculated in the equations
\begin{eqnarray}
 P_{\rm flow} \ &=& \ \sum_i M_{{\rm flow}, i} v_i, \\
 E_{\rm flow} \ &=& \ \frac{1}{2}\sum_i M_{{\rm flow}, i} v_i^2.
\end{eqnarray}
The derived $P_{\rm flow}$ and $E_{\rm flow}$ for the blueshifted lobe were 568.6\,$M_\odot\kms$ and $7.4\times10^{46}$\,ergs, respectively, while those of the redshifted lobe were 771.3\,$M_\odot\kms$ and $14.0\times10^{46}$\,ergs, respectively, as summarized in Table\,\ref{tab:1}.

The errors on the $M_{\rm flow}$ estimate mainly depend on the assumptions on $T_{\rm flow, ex}$ and $\tau_{\rm flow, 12}$. 
As $^{12}$CO spectra are affected by self-absorption (Figures\,\ref{fig:co}(b) and (c)), the derived $T_{\rm flow, ex}$ is possibly underestimated.
On the other hand, the assumption of the constant $\tau_{\rm flow, 12}$ makes $M_{\rm flow}$ slightly overestimated. 

The dynamical timescale $t_{\rm dyn}$ was estimated as 
\begin{equation}
t_{\rm dyn} \ = \ r_{\rm max}/v_{\rm max},
\end{equation}
where $r_{\rm max}$ is the maximum projected distance between the outflow lobe and the peak of the core \citep{dow2007}. 
The derived $t_{\rm dyn}$ were $2.0\times10^4$\,yrs and $1.1\times10^4$\,yrs for the blueshifted and redshifted lobes, respectively. 
As discussed in Section\,\ref{sec:res:2}, the present estimates of $t_{\rm dyn}$ give lower limits, as the spatial distribution of the outflow are hardly resolved as seen in Figure\,\ref{fig:co}(a). 
It is also possible that the outflow cavity is face-on and the estimated $r_{\rm max}$ accounts for only a small fraction of the real outflow lengths. 

\section{Estimates of the molecular mass and density in AGAL337.9-S and AGAL337.9-N}
The H$_2$ column density $N_{\rm core}$ and H$_{2}$ mass $M_{\rm core}$ of AGAL337.9-S were estimated using the following formulas from the 870\,$\mu$m map in Figure\,\ref{fig:ir}(e); 
\begin{eqnarray}
N_{\rm core} \ &=& \ \frac{R_{\rm g/d} F_{870}}{2.8 m_{\rm H} \kappa_{870} B_{870} (T_{\rm dust}) \Omega_{\rm beam}} \label{eq:ncore}\\
M_{\rm core}  \ &=& \ \frac{R_{\rm g/d} S_{870} D^2}{\kappa_{870} B_{870} (T_{\rm dust})}, \label{eq:mcore}
\end{eqnarray}
where $T_{\rm dust}$ is dust temperature, and $R_{\rm g/d}$ is the gas-to-dust ratio, where 100 is assumed. 
$F_{870}$ and $S_{870}$ are the flux density per beam in Jy\,beam$^{-1}$ and the integrated flux in Jy of the 870\,$\mu$m image, respectively.
In this analyses these values were estimated using the aperture photometries in Section \ref{sec:res:4}. 
$\kappa_{870}$ = 1.8\,cm$^{-2}$\,g$^{-1}$ is the dust opacity at 870\,$\mu$m \citep{oss1994}. 
$B_{870} (T_{\rm dust})$ is the Planck function for $T_{\rm dust}$. $\Omega_{\rm beam}$ is the beam solid angle of the ATLASGAL observations and was $9.817\times10^{-9}$\,str for a $19.2''$ beam.
By substituting $T_{\rm dust}$ of 28\,K, which was estimated from the fits for the SED in Section \ref{sec:res:4}, in the equations (\ref{eq:ncore}) and (\ref{eq:mcore}), $N_{\rm core}$ and $M_{\rm core}$ were calculated as summarized in Table\,\ref{tab:2}. 
The derived $N_{\rm core}$ was as high as $\sim$3.7$\times10^{23}$\,cm$^{-2}$ at the peak, while $M_{\rm core}$ was 1040\,$M_\odot$.
By assuming a sphere with the 2$\sigma$ radius of the core in AGAL337.9-S, which corresponds to 0.2\,--\,0.3\,pc as measured in Section \ref{sec:res:3}, the number density of the core $n_{\rm core}$ was derived as 1.3\,--\,4.5\,$\times10^4$\,cm$^{-3}$.

For comparison, we estimated $N_{\rm core}$ and $M_{\rm core}$ of AGAL337.9-S using the Mopra $^{13}$CO and C$^{18}$O data with an assumption of the local thermodynamic equilibrium (LTE).
The excitation temperature of the core $T_{\rm core, ex}$ was estimated from the $^{13}$CO data. 
We calculated the optical depth $\tau_{\rm core, 13}$ as,
\begin{eqnarray}
\frac{T_{\rm 13}}{T_{\rm 18}} = \frac{1-\exp(-\tau_{\rm core, 13})}{1-\exp{(-\tau_{\rm core, 13}/({\rm [^{13}O]/[^{18}O]})}}, \label{eq:1c}
\end{eqnarray}
where $T_{\rm 18}$ and $R_{13/18}$ is the observed line intensity of the C$^{18}$O emission.
$R_{13/18}$ can be calculated with the equations given by \citet{wil1994} as follows;
\begin{eqnarray}
\frac{\rm [^{16}O]/[^{18}O]}{\rm [^{12}C]/[^{13}C]}  = \frac{58.8 \times D_{\rm gc} + 37.1}{7.5\times D_{\rm gc}+7.6} \sim 7.4. \label{eq:1d}
\end{eqnarray}
The dervied $\tau_{\rm core, 13}$ and $\tau_{\rm 18}$ were $\sim$1.7 and $\sim$0.2, respectively, with $T_{\rm core, ex}$ = 30\,K derived using $T_{\rm 13}$ of 24\,K.
[$^{16}$O]/[$^{18}$O] was calculated as $58.8 \times D_{\rm gc} + 37.1 \sim 370$ \citep{wil1994}.
We here assumed that the gas and the dust are in thermal equilibrium, which is a reasonable assumption in the dense cores \citep[e.g.,][]{hil2005}.
The derived mass was $\sim$950\,$M_\odot$, which is almost the same with the estimate from the 870\,$\mu$m data (Table\,\ref{tab:2}).

The physical properties of AGAL337.9-N were estimated in the same way as were done for AGAL337.9-S.
The fit for the SED in Section \ref{sec:res:4} indicates $T_{\rm dust}$ of 26\,K in AGAL337.9-S.
The 2$\sigma$ radius of AGAL337.9-N was measured as 0.2\,pc.
As summarized in Table\,\ref{tab:2}, the estimated $M_{\rm core}$ of AGAL337.9-N ranges from 180\,$M_\odot$ for C$^{18}$O to 230\,$M_\odot$ for 870\,$\mu$m. 
$N_{\rm core}$ at the peak position, while $N_{\rm core}$ was 0.5\,--\,1.2$\times10^{23}$\,cm$^{-2}$.
Note that the present estimates in AGAL337.9-N with the 870\,$\mu$m image was affected from the contamination by RMS\,G337.9266-00.4588, and are may be slightly overestimated.

\allauthors

\listofchanges

\end{document}